\documentclass{ws-procs961x669}            

\usepackage{hyperref}
\begin{document}
\title{The Top Quark:\\ Past, Present, and Future}

\author{R. Sekhar Chivukula$^*$}

\address{Department of Physics and Astronomy, Michigan State University\\
East Lansing, MI 48824, USA\\
$^*$E-mail: sekhar@msu.edu\\
\url{http://www.pa.msu.edu/~sekhar}}

\begin{abstract}
In this talk I discuss the widespread impact of the top quark on phenomena in elementary particle physics as codified through the Standard Model (SM), its important role in motivating the possiblity of physics beyond the Standard Model (BSM), and its use as a signal for detailed studies of the SM and searches for BSM physics.
\end{abstract}


\bodymatter

\bigskip

\section{Past - Top Matters}\label{sec:past}

In the Standard Model the masses of all of the elementary particles (and, in particular, the ``current" masses of the quarks) arise through couplings between the particles and the Higgs boson.\footnote{This is true of the Higgs boson itself, whose mass arises through its {\it self-interactions}.}
With a mass of approximately 173 GeV/$c^2$, the top quark is the heaviest elementary particle discovered to date -- and therefore has the strongest (Yukawa) coupling to the Higgs boson of all known particles. Furthermore, the top quark mass violates two different kinds of symmetries that the Standard Model (SM) would possess were the quarks all massless  -- the ``custodial" $SU(2)$ symmetry \cite{Sikivie:1980hm} which insures that the weak-interaction $\rho$-parameter 
\begin{equation}
\rho=\frac{M^2_W}{M^2_Z \cos^2\theta_W},
\end{equation}
is equal to one at tree-level, and quark flavor symmetries \cite{Chivukula:1987py} which would prevent the appearance of flavor-changing neutral currents. 

The presence of non-degenerate and non-zero quark masses (and, in the case of the $\rho$-parameter, the hypercharge interactions) explain, through radiative corrections,  the observed deviations in the $\rho$-parameter from one and the presence of the flavor-changing neutral currents observed in neutral meson mixing. Because of its large coupling to the Higgs sector, the largest of these radiative corrections typically arise from loop processes involving the top quark. Hence, the presence of the top quark is essential to ability of the Standard Model to account for observed particle phenomenon.

\begin{figure}[h]
\begin{center}
\includegraphics[width=0.8\textwidth]{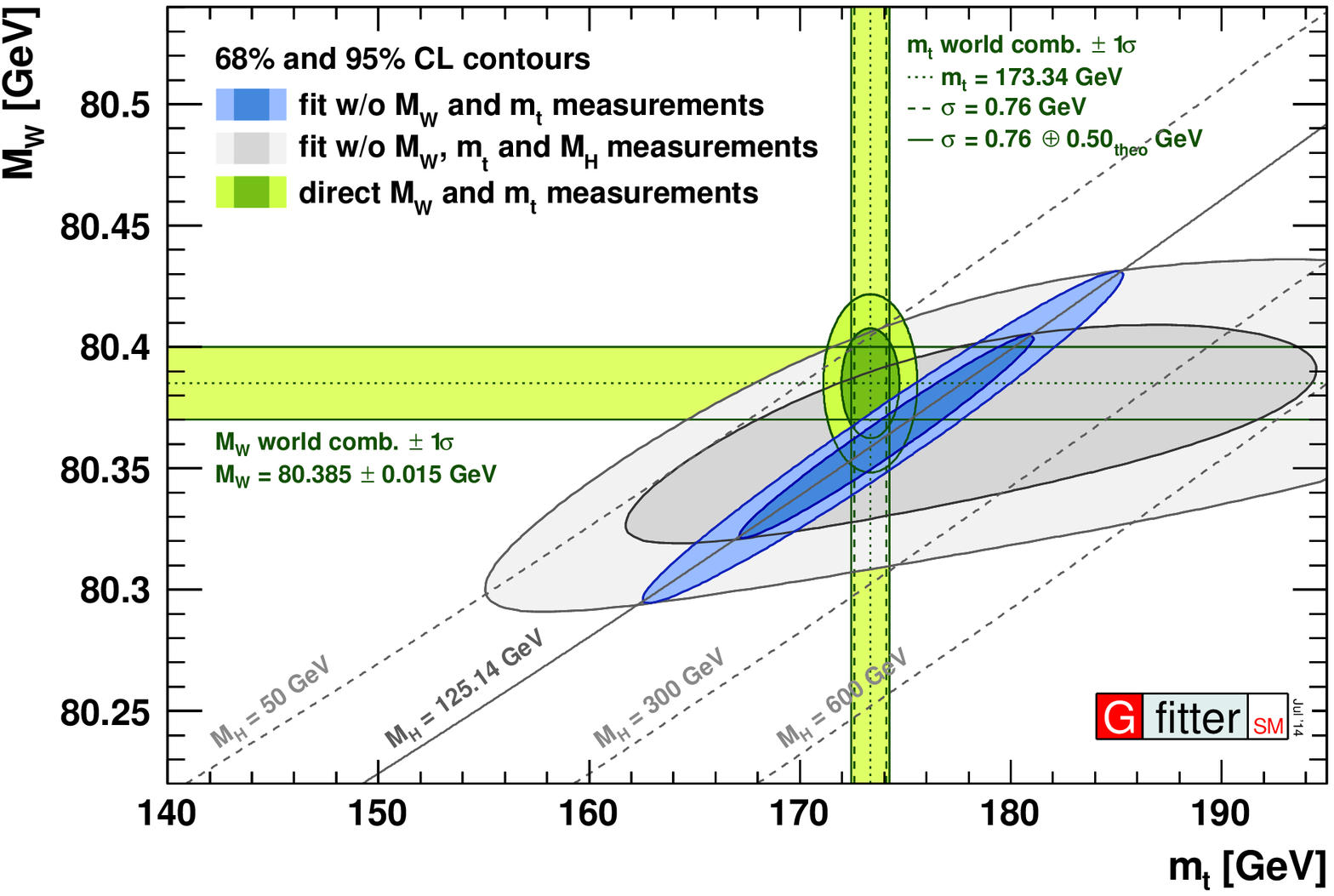}
\end{center}
\caption{Consistency\cite{Baak:2014ora} of the Standard Model. The influence of the top quark on precisely measured electroweak quantities is illustrated by the ``slope" of the diagonal, blue, fit-region -- showing the tight correlation needed between $M_W$ and $m_t$ to be consistent with experimental results.}
\label{fig:one}
\end{figure}

The consistency of the Standard Model is illustrated\cite{Baak:2014ora} in  Fig. \ref{fig:one}, by the agreement of direct measurements of the top quark and $W$ boson masses (green horizontal and vertical regions) and the best-fit indirect determination of these masses obtained by using all other electroweak data (blue, oval, diagonal regions). The influence of the top quark is illustrated by the ``slope" of the diagonal blue region -- demonstrating how a change in the top quark mass would only be consistent with experimental data by changing the $W$ boson mass as well.\footnote{Indeed, prior to the discovery of the Higgs boson, the measured values of $M_W$ and $m_t$ favored a Higgs mass in the 100 GeV range.}

The effect of the top quark on meson mixing is illustrated through the properties of $B_d - \bar{B}_d$ mixing. The dominant contribution to this process from the Standard 
Model comes through the box diagram,\footnote{Additional ``crossed" diagrams are not shown, but have the same dependence on the top quark.} illustrated below,
\begin{equation}
{\raisebox{-17pt}{\includegraphics[width=1.5truein]{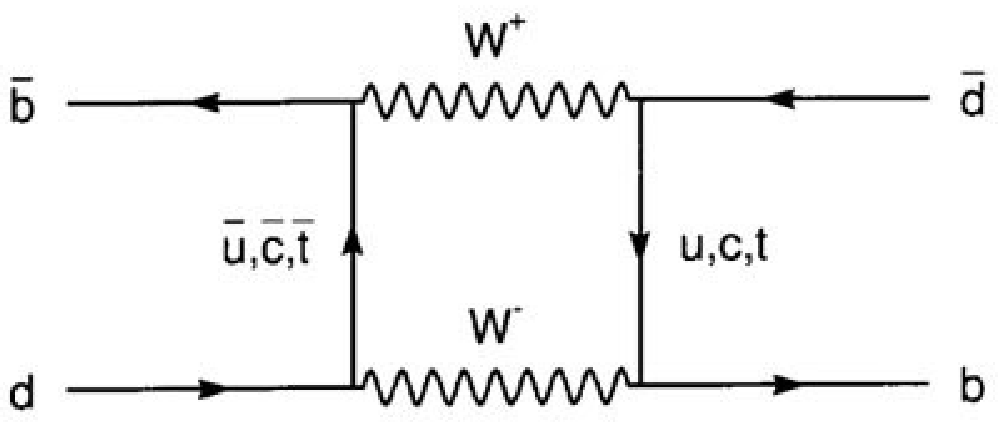}}}~,
\end{equation}
which leads to a prediction for the splitting of $B$-meson CP eigenstates (and, hence, to the oscillation frequency of flavor eignenstates) of order
\begin{equation}
\Delta m_B \propto m^2_t |V_{tb} V^*_{td}|^2 \approx 0.00002 \left(\frac{m_t}{{\rm GeV}/c^2}\right)^2\,{\rm ps}^{-1}~,
\end{equation}
where $V_{tb,td}$ are the relevant CKM quark-mixing elements.
The initial observation of $B_d$-meson mixing by the ARGUS collaboration\cite{Prentice:1987ap} found $\Delta m_B \simeq 0.5$ ps$^{-1}$, and therefore concluded (on the basis of what was then known about the quark-mixing matrix) that $m_t > 50$ GeV/$c^2$ -- the first indication of a heavy top quark!

In the cases above, the top quark Yukawa coupling enters via the coupling of the massive $W$ boson, whose longitudinal component is (gauge-equivalent to) an ``eaten" charged Goldstone boson from the Higgs doublet. The top quark coupling directly to the neutral Higgs boson itself, on the other hand, leads to the dominant production mechanism for the Higgs boson at hadron colliders,
\begin{equation}
{\raisebox{-17pt}{\includegraphics[width=1.25truein]{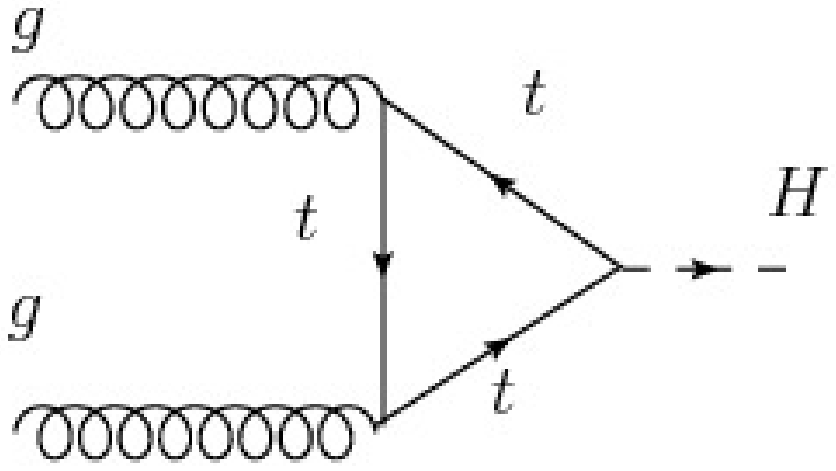}}}~,
\end{equation}
production via gluon fusion,\cite{Georgi:1977gs} which occurs through a virtual top quark loop.

\begin{figure}[h]
\begin{center}
\includegraphics[width=0.8\textwidth]{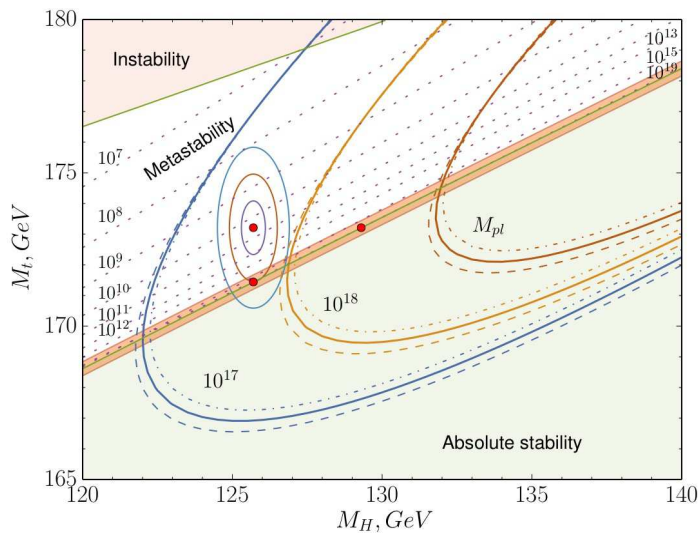}
\end{center}
\caption{Properties of the (two-)loop corrected Higgs potential in the Standard Model.\cite{Bednyakov:2015sca} Note that, for the central current values of top quark and Higgs boson masses, the Standard Model electroweak potential is metastable, a property which is exquisitely sensitive to the top quark mass.}
\label{fig:two}
\end{figure}

Finally, the large top quark Yukawa coupling alters -- at loop-level -- the properties of the electroweak potential. As shown\cite{Bednyakov:2015sca} in Fig. \ref{fig:two}, the Standard Model electroweak potential is metastable. This property depends very senstively on the top quark mass, with the potential becoming stable if the top quark had a mass of around 170 GeV/$c^2$ or lower. 

These examples show that {\it the top quark matters} -- in fact, the agreement of the Standard Model with observed particle phenomenology occurs only because of the virtual contributions arising from top quark exchange.

\section{Present - Top Couplings}\label{sec:present}

\begin{figure}[h]
\begin{center}
\includegraphics[width=0.8\textwidth]{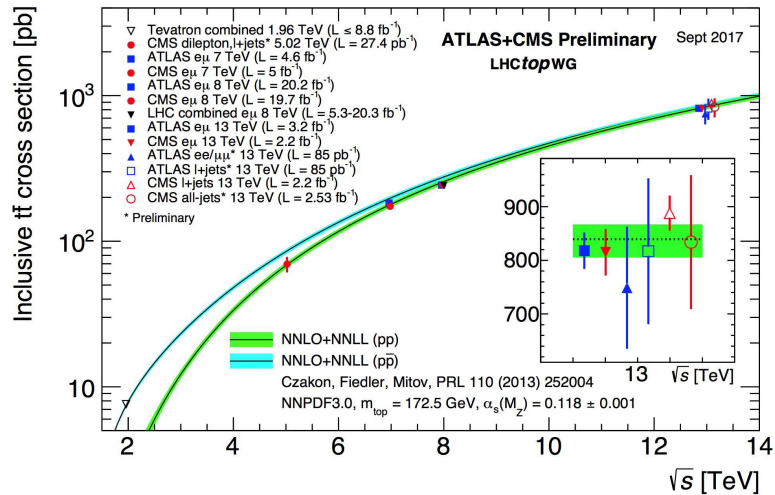}
\end{center}
\caption{Summary\cite{ttbarxsection} of top-pair production cross section measurements from the Tevatron and the LHC. The agreement between measurement and theory is a confirmation that the top quark is a color-triplet. }
\label{fig:three}
\end{figure}

The LHC, and the Tevatron before it, perform as veritable ``top-factories" -- allowing for detailed measurements of top quark couplings. For example, Fig. \ref{fig:three} summarizes\cite{ttbarxsection} the cross section measurements for top-pair production at the LHC and Tevatron at various energies. The close agreement between measurement and the theoretical QCD cross section serves as a sensitive check that the top quark is a color-triplet quark, just like the light quarks.

\begin{figure}[h]
\begin{center}
\includegraphics[width=0.8\textwidth]{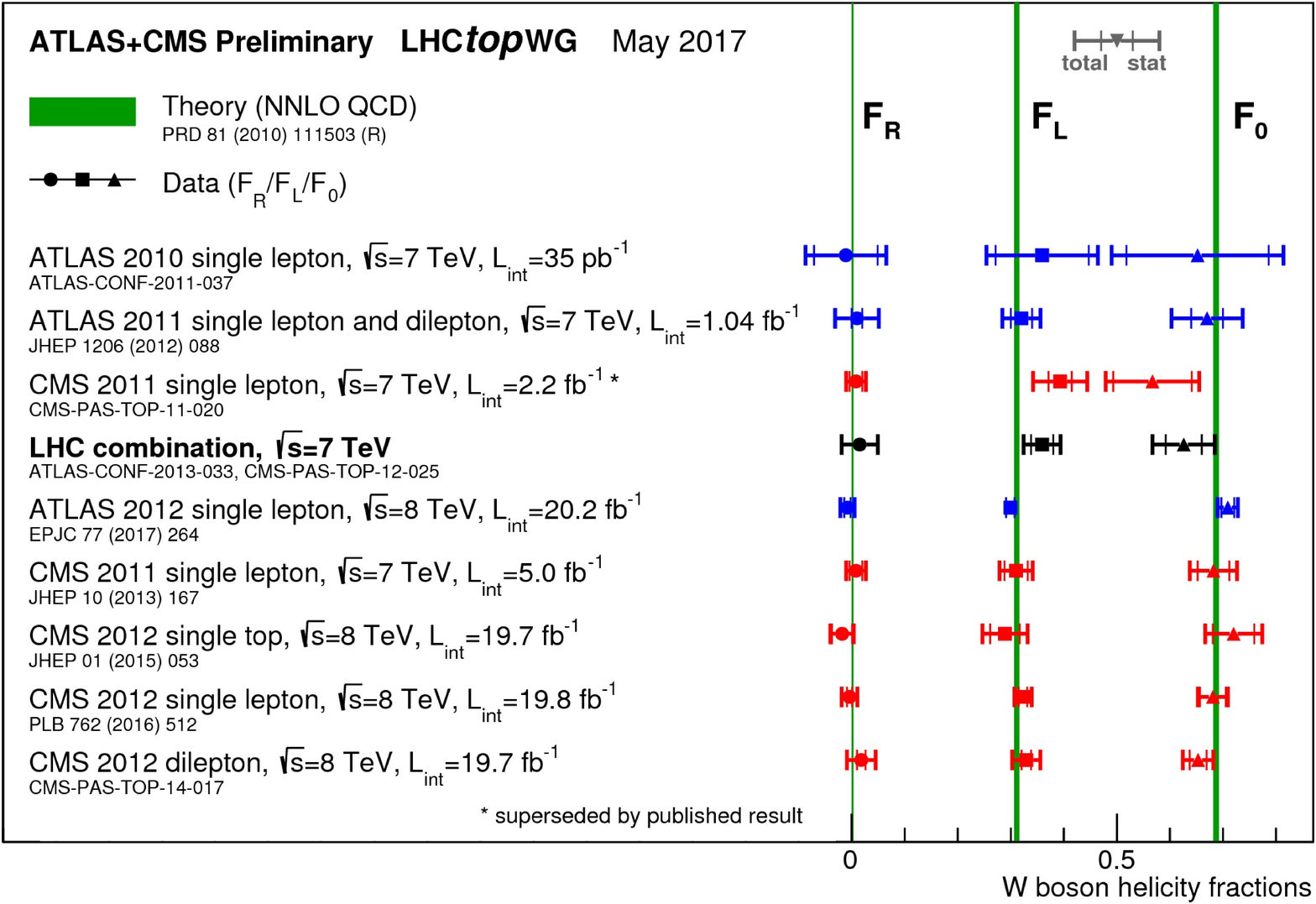}
\end{center}
\caption{Summary\cite{whelicity} of measured W helicity fractions by ATLAS and CMS at 7 and 8 TeV, compared to the respective theory predictions. The results are consistent with the ``V-A" structure of the weak-interactions, and the large fraction of longitudinally polarized $W$ bosons serves as an indirect check of the top quark Yukawa coupling to the Higgs doublet. }
\label{fig:four}
\end{figure}

Similarly, initial direct measurements of the top quark width\cite{CMS:2016hdd,ATLAS:2017faa} are consistent with the top quark coupling to the $W$ boson with the strength predicted by the Standard Model. More incisively, Fig. \ref{fig:four} shows that the helicity of the $W$ boson produced in the decay $t \to Wb$ is consistent with the standard ``V-A" structure of the weak coupling. Note, in particular, that the roughly 40\% longitudinal $W$ boson fraction in this decay is again a measure of the top quark's large Yukawa coupling.

\begin{figure}[h]
\begin{center}
\includegraphics[width=0.5\textwidth]{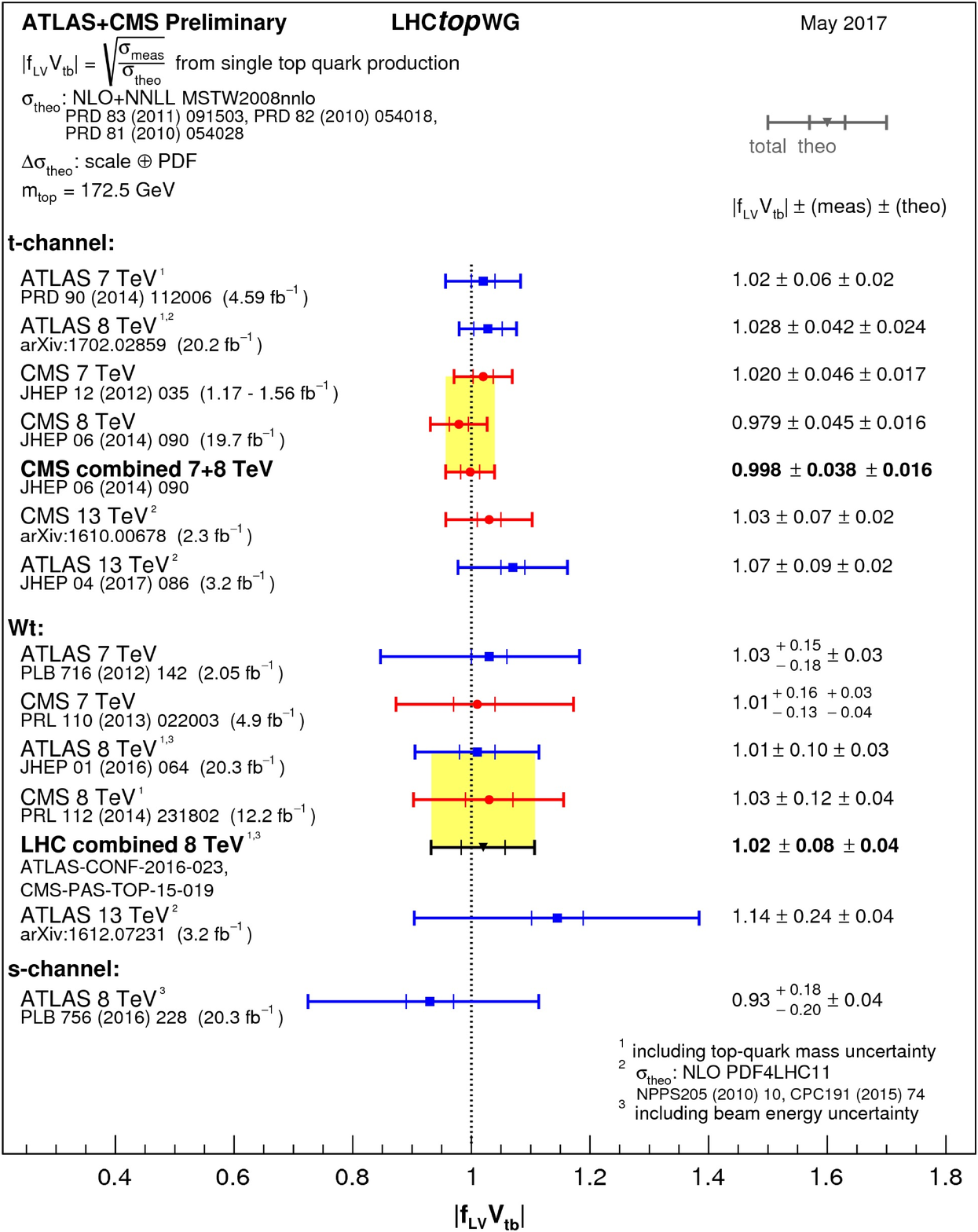}
\end{center}
\caption{Summary\cite{singletop} of the ATLAS and CMS extractions of the CKM matrix element $V_{tb}$ from single top quark production measurements. }
\label{fig:five}
\end{figure}

Furthermore, as summarized in Fig. \ref{fig:five}, measurements of the single-top production cross section (through various channels) yields a measurement of the CKM matrix element $V_{tb}$ -- which is very close to one.

\begin{figure}[h]
\begin{center}
\includegraphics[width=0.6\textwidth]{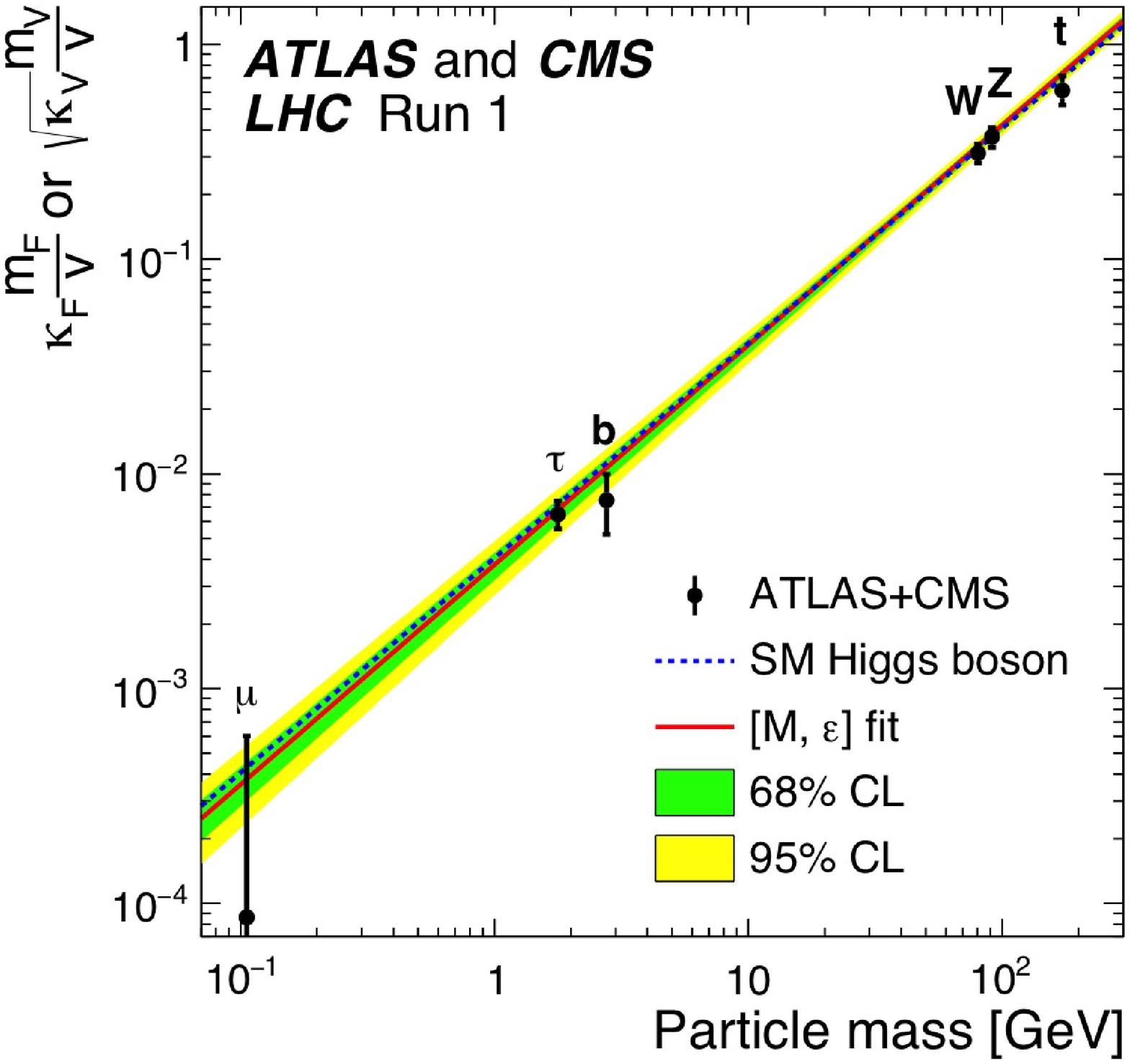}
\end{center}
\caption{Fit to Higgs boson couplings based on a combined Run 1 ATLAS and CMS analysis.\cite{Khachatryan:2016vau}  }
\label{fig:six}
\end{figure}

As noted in the previous section, the top quark's large Yukawa coupling makes the gluon-fusion cross section the dominant production mode for the Higgs boson. Using the observed properties of the Higgs boson we can, in the context of the Standard Model, extract measurements (or, in some cases, limits) on the coupling of the Higgs to various particles. Current measurements are summarized in Fig. \ref{fig:six}, which again demonstrates the consistency of the Standard Model, and graphically shows that the Higgs boson couples most strongly to the top quark.

\begin{figure}[h]
\begin{center}
\includegraphics[width=0.6\textwidth]{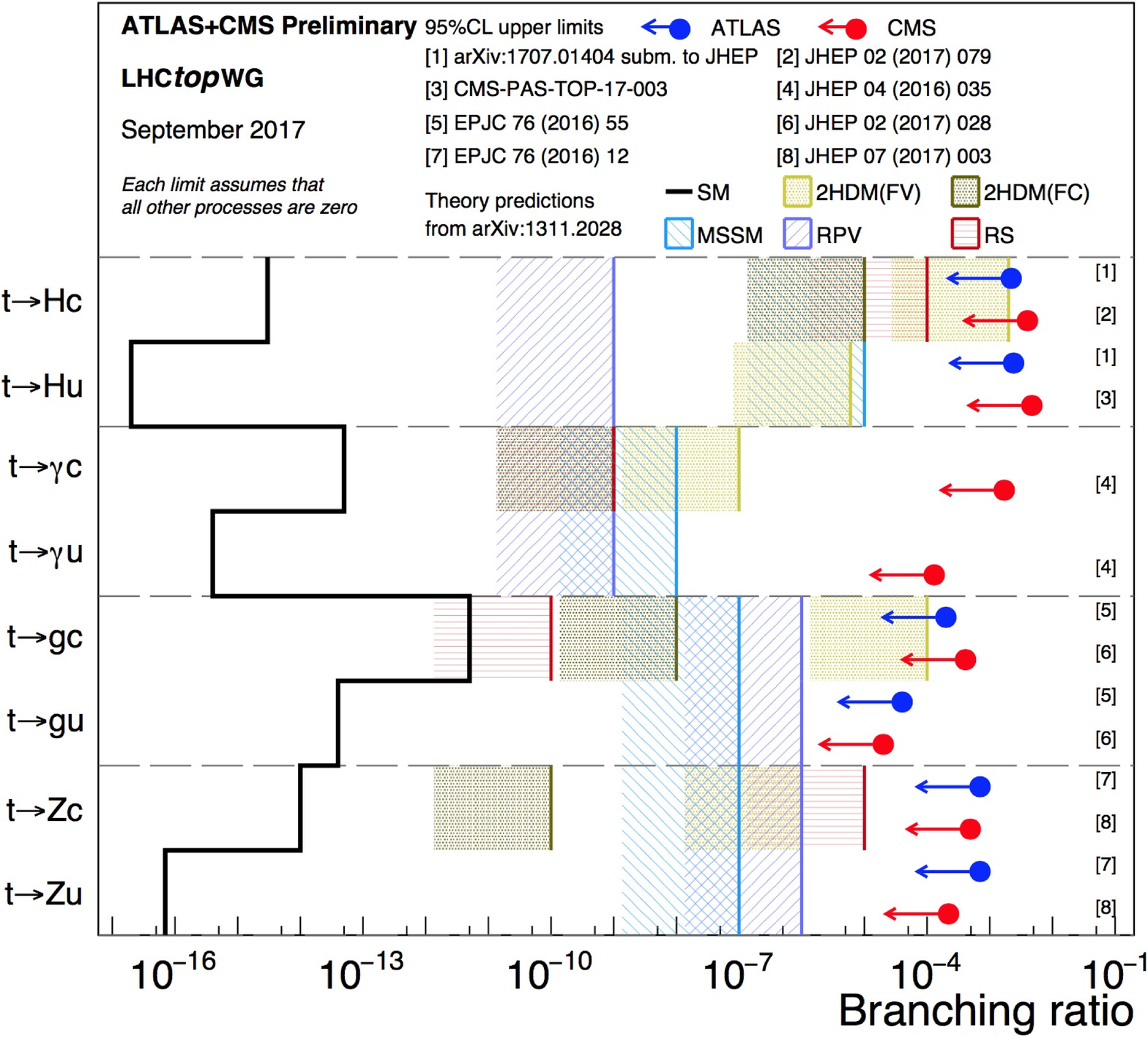}
\end{center}
\caption{Summary\cite{flavorchangintop} of the 95\% confidence level observed limits on the branching ratios of the top quark decays via flavour changing neutral currents to a quark and a neutral boson $t\to Xq$ ($X=g$, $Z$, $\gamma$ or $H$; $q=u$ or $c$) by the ATLAS and CMS Collaborations compared to several new physics models.  }
\label{fig:seven}
\end{figure}

With the large number of top quark pairs produced at the LHC, ATLAS and CMS can now begin to set meaningful bounds on flavor-changing top quark decays. In Fig. \ref{fig:seven} we see a summary of bounds on decays of the form $t \to X + q$, where
$X=g,\, Z,\, \gamma$ or $H$, and $q=u$ or $c$. We see that, while the bounds are not near the Standard Model predictions, they are beginning to be sensitive to various models of new physics.

{\it Presently}, we are beginning to study top quark properties in detail and with precision, and to set limits on
new physics from the results.

\section{Future - Naturalness and the Hierarchy Problem}\label{sec:futurei}

As foreshadowed by our discussion of the effect of the top quark on vacuum stability illustrated in Fig. \ref{fig:two}, the top quark strongly affects the electroweak symmetry breaking sector (EWSB). The most direct effect of the top on EWSB is in terms of the ``little hierarchy problem".\cite{Barbieri:2000gf} As illustrated in the diagram below, the contribution from top quark loops to the dimension-two Higgs mass term, calculated using a ``cutoff" scale $\Lambda$, is
\begin{equation}
{\raisebox{-10pt}{\includegraphics[width=1.25truein]{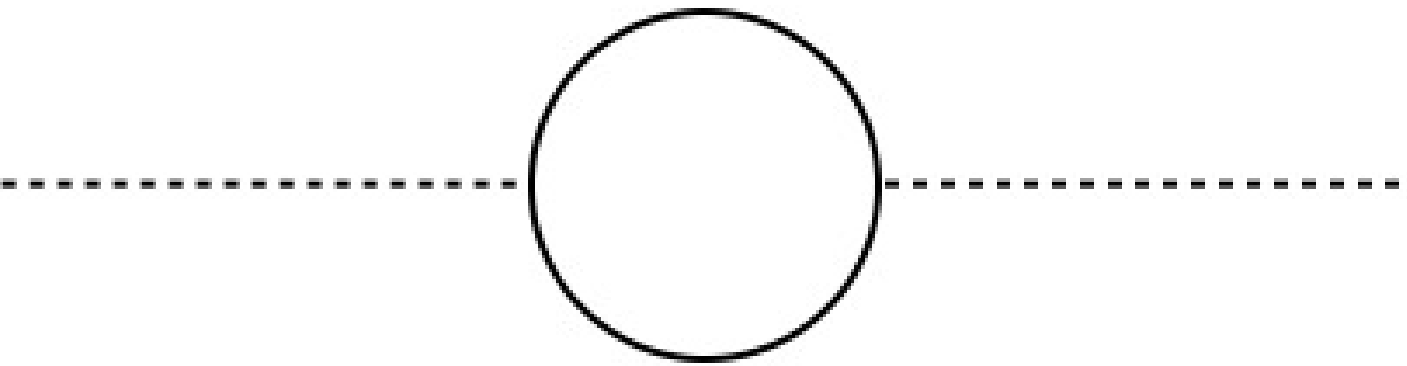}}} \Rightarrow\, \delta m^2_H \propto \frac{2\lambda^2_t \Lambda^2}{16 \pi^2}~.
\label{eq:mhsq}
\end{equation}
By interpreting $\Lambda$ as the scale of physics underlying the Standard Model, and requiring that $\delta m^2_H < (100\,{\rm GeV})^2$, we see that the scale of new physics cannot be larger than one or a few TeV. The Standard Model is therefore {\it unnatural} -- it is, generically, very sensitive to any new physics occuring at scales of order a TeV or higher. 

Since the Standard Model does not encompass gravity, we know that it is a low-energy effective theory description which must break down at some higher scale -- presumably at a scale lower than the Planck mass, $M_{pl} = {\cal O}(10^{16}\,{\rm TeV})$.  Therefore the low-energy Standard Model effective theory suffers from the {\it hiearchy} problem:  {\it if} there is no new physics between the electroweak scale and the Planck scale, then the properties of that underlying theory must be exquisitely {\it adjusted} (``tuned") in order for the Higgs mass, and hence the electroweak scale, to be of order 1 TeV. For example, if the scale of new physics is indeed of order the Planck Scale, we must explain how $m^2_H$ arises in the low-energy Standard Model (effective theory) and is of order $10^{-32}$ times smaller!

The naturalness and hierarchy problems motivate the search for new physics beyond the Standard Model at the TeV scale. While it is, in principle, possible that no new physics occurs until very high energies, and that the tuning required to keep the weak scale of order a TeV occurs by {\it accident}, it is important to explore the possiblity that the TeV scale arises dynamically.\footnote{For example, we are not concerned with the occurrence of the QCD scale of order a GeV since this scale occurs via dimensional transmutation and the asymptotic freedom of non-abelian gauge-theories.\cite{Coleman:1973jx,Politzer:1973fx,Gross:1973id}} One possiblity is that the elecroweak scale is associated with the dynamical breaking of a symmetry that would, if unbroken, leave the Higgs boson massless in the low-enery effective theory. The dynamically breaking of that symmetry would ``generate" EWSB scale and, typically, give rise to new states at the TeV scale. The consistency of the low-energy theory would then be reflected in a ``cancellation" of contributions to the Higgs boson mass -- eliminating the quadratic divergence implicit in the diagram in Eq. \ref{eq:mhsq}.

\begin{figure}[h]
\begin{center}
\includegraphics[width=0.95\textwidth]{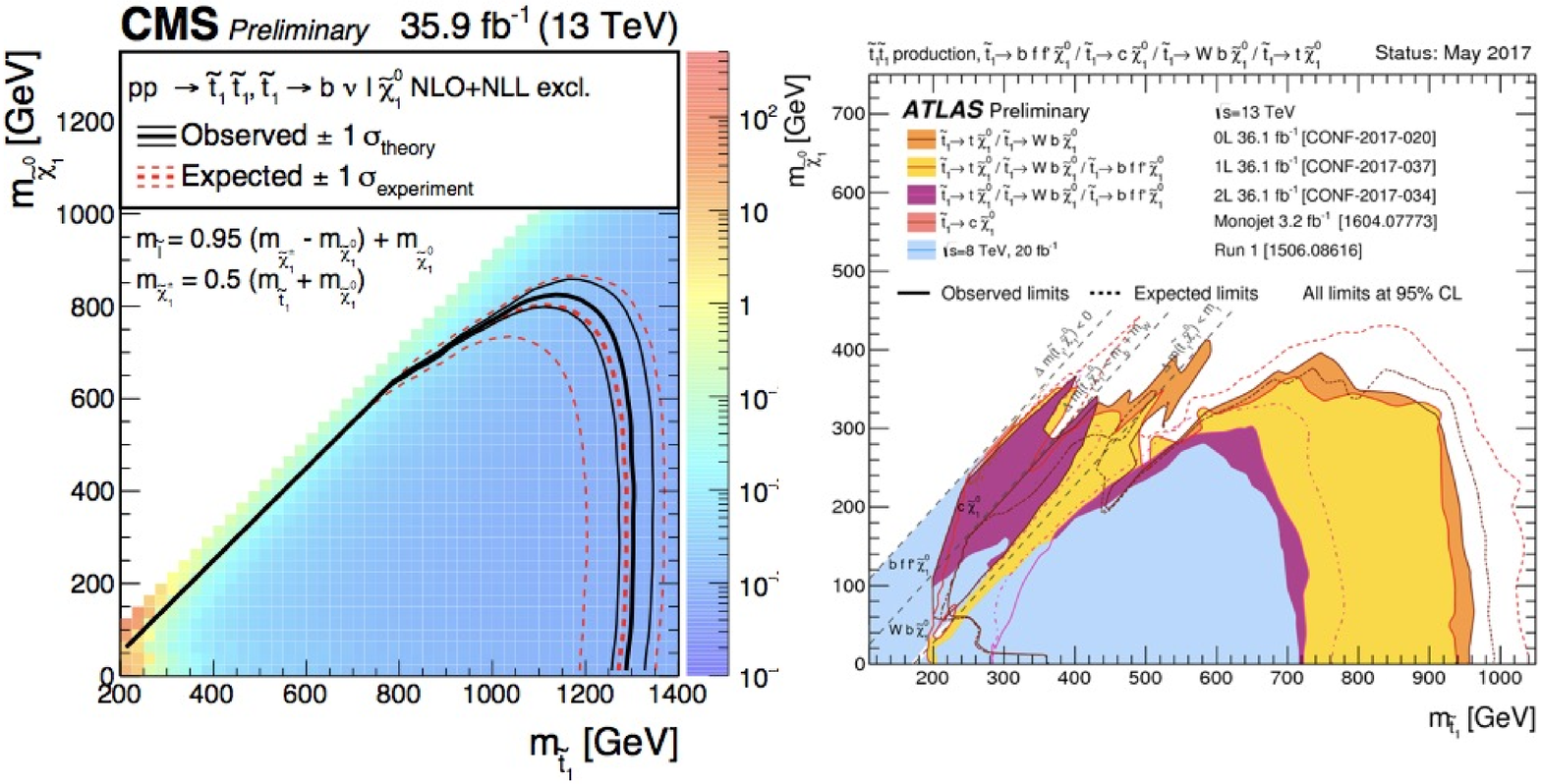}
\end{center}
\caption{Top squark searches. CMS left\cite{Sirunyan:2017leh} and ATLAS right\cite{ATLAS-topsquark}.  }
\label{fig:eight}
\end{figure}

The most popular new symmetry introduced to protect the electroweak scale is supersymmetry.\footnote{See review articles in \cite{Patrignani:2016xqp}, and references therein.} Supersymmetry pairs bosonic and fermionic states, and therefore the Higgs mass term(s) become related to the chiral symmetry which would be present if the corresponding fermionic states were massless. Diagramatically, for the top quark contribution in Eq. \ref{eq:mhsq}, there are contributions from the squarks ($\widetilde{t}_{R,L}$), the scalar partners of the top quark
\begin{equation}
\raisebox{-32pt}{\includegraphics[width=1.25truein]{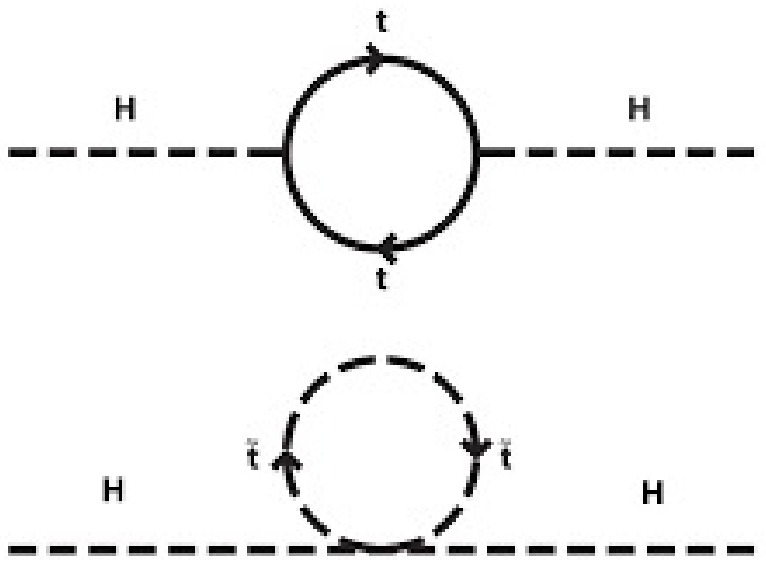}} \Rightarrow \delta m^2_H \propto m^2_{\widetilde{t}}~.
\end{equation}
As noted, the residual contributions to the Higgs boson mass are proportional to the 
masses of the SUSY partners in the loop (assuming $m_{\widetilde{t}} \gg m_t$) -- and hence, to avoid the hierarchy problem,  we would expect new supersymmetric partner states in the TeV mass range. Of particular note, however, is that supersymmetry protects the Higgs boson mass from potentially larger contributions from higher scale physics ({\it e.g.} quantum gravity) so long as those additional sectors are supersymmetric. Fig. \ref{fig:eight} summarizes recent searches for top-squarks -- and, although now beginning to be sensitive to top-squarks in the TeV mass range, there is as yet no sign of the top-squark.

An alternative to supersymmetry is to construct a theory in which the Higgs boson is itself a Goldstone boson of a spontaneously broken global symmetry.\cite{Kaplan:1983fs} In this case, the Higgs mass-term is protected by a ``shift symmetry" in the low-energy effective lagrangian.\cite{Coleman:1969sm} The global symmetries generally used involve chiral symmetries of fermions which feel a new (so-far undiscovered) strong interaction, which dynamically breaks this chiral symmetry in a way analogous to chiral symmetry breaking in QCD. The Higgs is then a Goldstone boson bound-state of these underlying fermionic degrees of freedom, analogous to how the pions of QCD are bound states of the quarks. This scenario therefore realizes the possiblity that the Higgs boson is {\it not} a fundamental scalar, but rather a bound state of fermions interacting via some strong (gauge theory) dynamics.  

\begin{figure}[h]
\begin{center}
\includegraphics[width=0.95\textwidth]{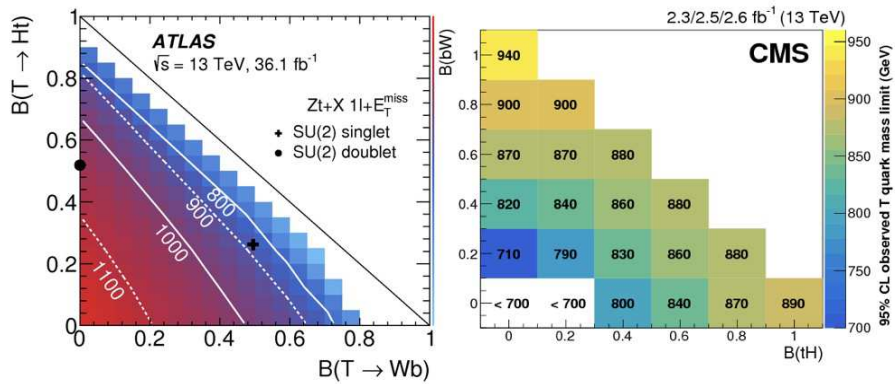}
\end{center}
\caption{Vector-quark searches. ATLAS left\cite{Aaboud:2017qpr} and CMS right\cite{Sirunyan:2017usq}.  }
\label{fig:nine}
\end{figure}

Small global symmetry-breaking interactions are then added to produce the Higgs boson mass terms and to produce the ``vacuum misalignment" necessary to break the electroweak symmetry. Among the symmetry-breaking terms needed are flavor-dependent ones. The Yukawa couplings necessary to generate the top quark mass, in particular, explicitly break the shift-symmetry, and the flavor symmetries must then be extended to the new fermionic degrees of freedom which are the consitutents of the Higgs boson. Diagramatically, just as in supersymmetry, additional TeV-scale states arise which cancel the potentially dangerous quadratically divergent top-loop contributions to the Higgs boson mass shown in Eq. \ref{eq:mhsq}. In this case, the additional states are typically ``vectorial" partners of the top quark (and potentially of the other fermions as well), new heavy color-triplet fermions whose right- and left-handed components both couple to the weak interaction $SU(2)$. Interestingly, in the low-energy theory, the diagrams involving the vector-like partners of top ($\chi$) typically interact with the Higgs through a pair-wise dimension-five interaction
\begin{equation}
{\raisebox{-20pt}{\includegraphics[width=2.0truein]{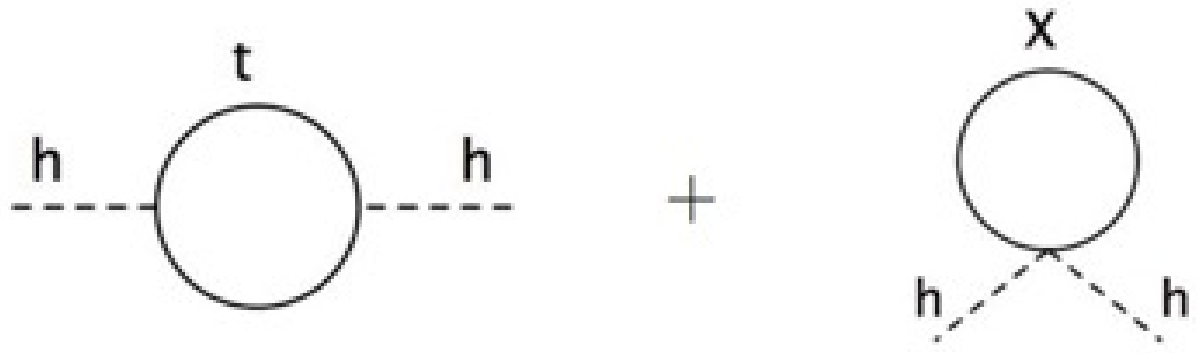}}} \Rightarrow \delta m^2_H \propto m^2_\chi~,
\end{equation}
whose strength is related via the global symmetries of the theory to the top quark Yukawa coupling.
Limits from LHC searches for vector-like top quark partners are illustrated in Fig. \ref{fig:nine} and, as in the case
of supersymmetry, although we are becoming sensitive to these states in the TeV region there is no evidence for these additional states.

The absence (so far) of new TeV-scale states described above suggests that perhaps we need to consider new kinds of physics beyond the Standard Model. One intriguing possiblity is ``neutral naturalness."\cite{Chacko:2005pe,Burdman:2006tz} The motivation for these models is based on observation that the additional TeV states above (the top squark or vectorial top-partners) are copiously produced at the LHC because they are colored. In addition, they are colored because the additional global symmetry introduced to protect the Higgs boson mass commutes with QCD -- hence the states related to the top must themselves be color triplets. Instead, if the global symmetries introduced include QCD (or some symmetry related to QCD) as a subgroup then, potentially,
the states related to the top quark that are introduced to cancel the quadratically divergent contributions to the Higgs boson mass ({\it e.g.} Eq. \ref{eq:mhsq}) can be color neutral. The phenomenology of these models is quite different, including potentially long-lived particles and exotic signals, and they are far less constrained.\cite{Curtin:2015bka}

For the {\it future}, the presence of the top quark -- and, in particular, its large Yukawa-coupling to the Higgs boson -- is a primary motivation for the presence of new physics at the TeV scale. The LHC was specifically built to explore the TeV region, and time will tell if the kinds of theories we discuss here are realized in nature.

\section{Future - Top as a Signal}\label{sec:futureii}

It is often said that ``yesterday's signal is tomorrow's background." With the ability top to be (pair) produced via the strong interactions, and its decay to $Wb$, the top quark is an important background in many searches for physics beyond the Standard Model. Given the fact that the top quark couples strongly to the EWSB sector, however, {\it the top is also an important part of the signal in many kinds of standard and non-standard physics.} Some examples are given here.

\begin{figure}[h]
\begin{center}
\includegraphics[width=0.95\textwidth]{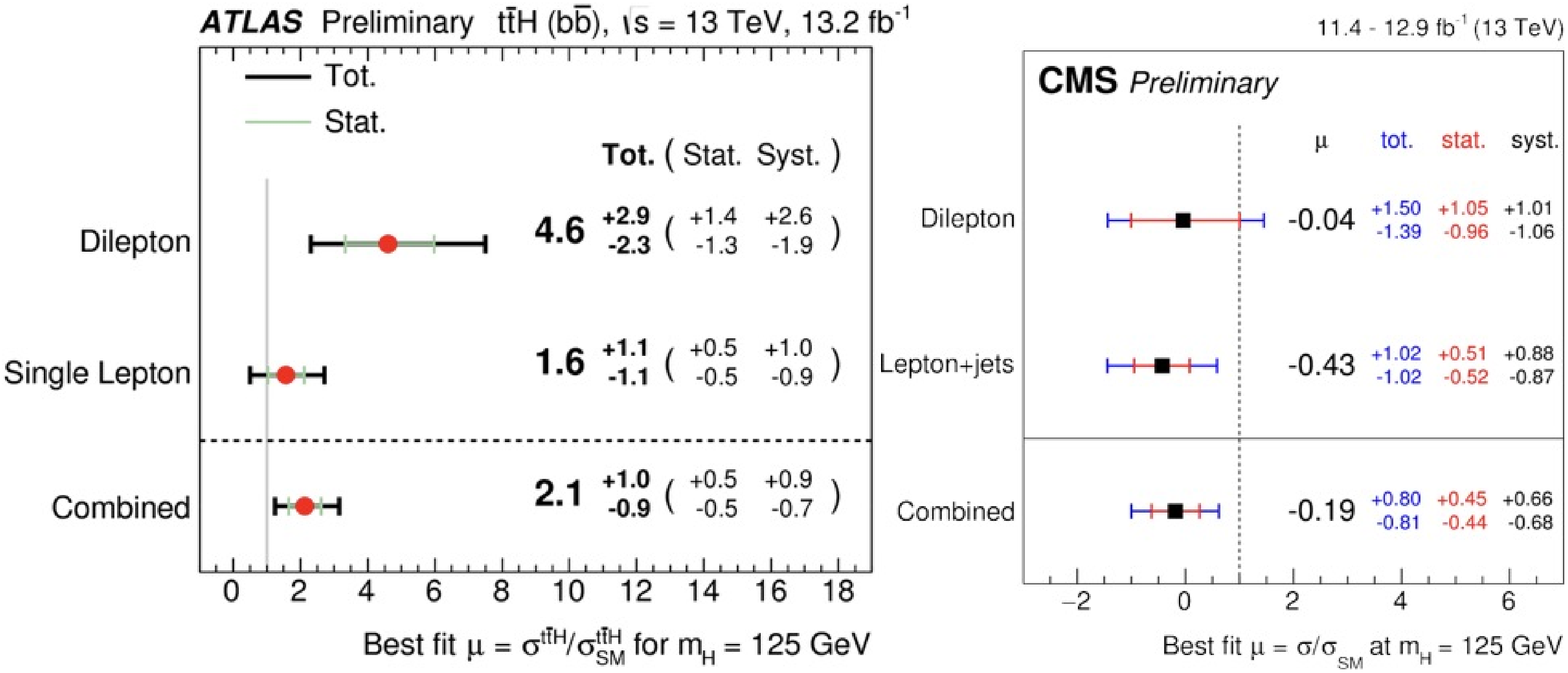}
\end{center}
\caption{ttH searches. ATLAS left\cite{ATLAS-CONF-2016-080} and ATLAS right\cite{CMS:2016vqb}.  }
\label{fig:ten}
\end{figure}

First, one crucial signal involving the top quark in the Standard Model is the in $t\bar{t}H$ production,
\begin{equation}
{\raisebox{-17pt}{\includegraphics[width=1.25truein]{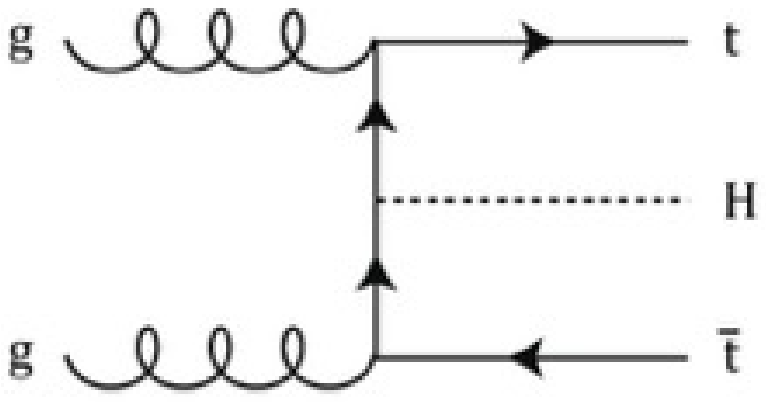}}}~,
\end{equation}
a process which would yield a direct measurement of the top quark Yukawa coupling. Current limits on this
process are shown in Fig. \ref{fig:ten}. As we see there, the measurements are only beginning to be sensitive to
this process -- but with additional data, the signal should become clearly visible, and we will be able to directly
test whether the Standard Model mass-production mechanism applies in the case of the top quark.\footnote{Similarly, observations
of the Higgs decay $H\to Z Z^*\to 4\ell$ gives us direct evidence of the Standard Model mechanism for generating the masses of
the electroweak gauge bosons.} 

\begin{figure}[h]
\begin{center}
\includegraphics[width=0.95\textwidth]{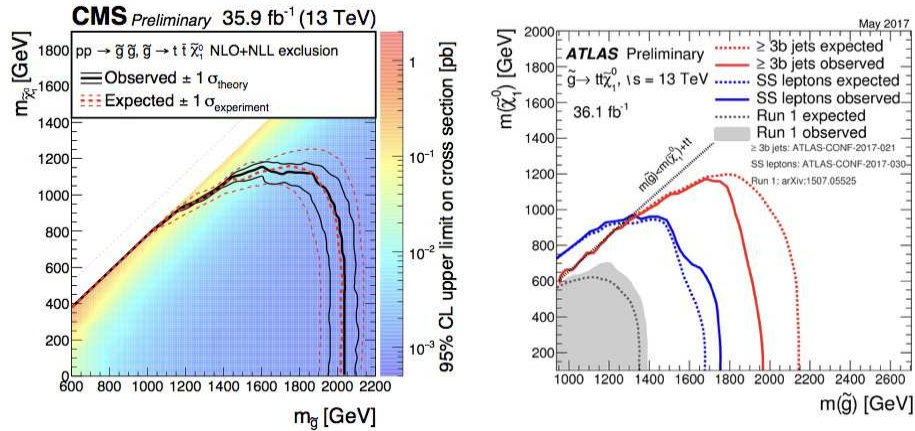}
\end{center}
\caption{Searches for pair-production of gluinos, which subsequently decay to top plus neutralino. CMS left\cite{CMS-PAS-SUS-16-050} and ATLAS right\cite{ATLAS-gluinotop}.  }
\label{fig:eleven}
\end{figure}

Second, the important contribution of the top-squark to protecting the Higgs boson mass in SUSY theories suggests that the top-squark may be substantially lighter than the scalar partners of the other quarks.\footnote{Just as quark mass-splittings and mixings give rise to flavor-changing neutral-currents, squark mass-splittings and mixings can as well ... so a theory with a large hierarchy of masses between the top-squark and the other scalar partners will need to have relatively small mixing between the third-generation of squarks with the other two in order to avoid dangerously large flavor-changing neutral currents.} This possiblity, along with the large gluino (the fermionic partner of the gluon) pair-production cross section at the LHC, motivate the search shown in Fig. \ref{fig:eleven} -- where the gluino decays to a top quark and the (assumed neutral and stable) lightest supersymmetric partner (LSP, $\widetilde{\chi}^0$). Displayed in the $(m_{\widetilde{g}},m_{\widetilde{\chi}^0})$ plane, we see that current limits extend to gluino masses of order 2 TeV, for LSP masses up to 1200 GeV.

\begin{figure}[h]
\begin{center}
\includegraphics[width=0.95\textwidth]{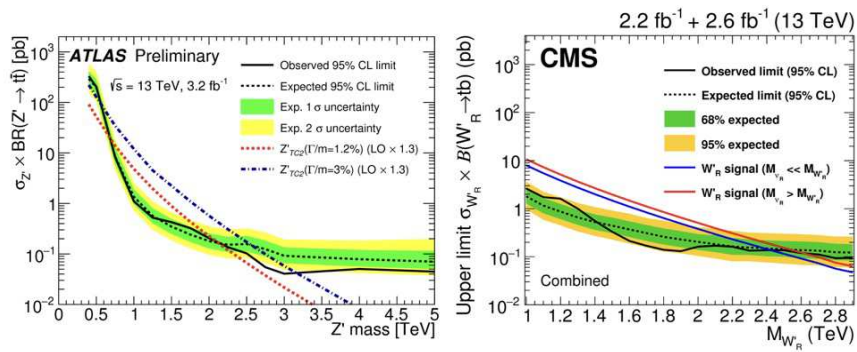}
\end{center}
\caption{Searches for resonances decaying to top. ATLAS left\cite{ATLAS-CONF-2016-014} and CMS right\cite{Sirunyan:2017ukk}.  }
\label{fig:twelve}
\end{figure}

Finally, the large coupling of the top quark to the EWSB sector suggests that the top quark could play a special role in theories where the elecroweak sector is extended. Considerations of this kind motivate the search for $W'$ and $Z'$ decaying to $tb$ and $tt$ as shown in Fig. \ref{fig:twelve}. We see here that, though model-dependent, these limits are now sensitive to new gauge-bosons decaying preferentially to third-generation quarks up to 2-3 TeV.

In short, {\it the top quark provides an important signal of new physics for the LHC now, and for the future.}

\section{Conclusions}

To summarize, to a very good approximation, we know that the top quark
\begin{itemize}
\item is a $(3,2)_{+1/6}$ (left) and $(3,1)_{+2/3}$ (right) under the $SU(3)_C \times SU(2)_W \times U(1)_Y$ guage interactions of the SM,
\item mixes relatively little with the other quark-generations, with $V_{tb}\simeq 1$, $|V_{ts}| \simeq 0.04$, and $|V_{td}| \simeq 0.009$,
\item and has a large Yukawa coupling, $y_t = \sqrt{2}m_t/v \simeq 1$.
\end{itemize}
The top quark casts a large shadow over particle phenomenology, generating the largest electroweak and flavor radiative-corrections, providing the dominant production mechanism for the Higgs boson at the LHC, motivating the presence of TeV scale new-physics, and playing an important role in signals for physics both in the Standard Model and beyond. 

Experiments at the LHC are now directly sensitive to new physics at the TeV scale, and the top quark will play a central role in experimental and theoretical advances in particle physics for the foreseeable future.

\section*{Acknowledgements}
I would like to thank the conference organizers for their hospitality, and Dennis
Foren for comments on the manuscript. This material is based in part upon work supported by the National Science Foundation under Grant No. 1519045.

\bibliographystyle{ws-procs961x669}
\bibliography{LP2017}

\end{document}